# New Fe-based superconductors: properties relevant for applications

M Putti, I Pallecchi, E Bellingeri, M Tropeano, C Ferdeghini, A Palenzona,
CNR-INFM-LAMIA and Università di Genova, Via Dodecaneso 33, I-16146 Genoa, Italy
C Tarantini, A Yamamoto, J Jiang, J Jaroszynski , F Kametani, D Abraimov, A Polyanskii, J D Weiss, E E Hellstrom, A Gurevich, D C Larbalestier
Applied Superconductivity Center, National High Magnetic Field Laboratory, Florida State University, 2031 East Paul Dirac Drive, Tallahassee, FL 32310, USA
R Jin, B C Sales, A S Sefat, M A McGuire, D Mandrus
Materials Science & Technology Division, Oak Ridge National Laboratory, Oak Ridge, TN 37831, USA
P Cheng, Y Jia, H H Wen
Institute of Physics, Chinese Academy of Sciences, Beijing 100190, People's Republic of China
S Lee, C B Eom
Department of Materials Science and Engineering, University of Wisconsin, Madison, WI 53706, USA

**Abstract**
Less than two years after the discovery of high temperature superconductivity in oxypnictide LaFeAs(O,F) several families of superconductors based on Fe layers (1111, 122, 11, 111) are available. They share several characteristics with cuprate superconductors that compromise easy applications, such as the layered structure, the small coherence length, and unconventional pairing, On the other hand the Fe-based superconductors have metallic parent compounds, and their electronic anisotropy is generally smaller and does not strongly depend on the level of doping, the supposed order parameter symmetry is s wave, thus in principle not so detrimental to current transmission across grain boundaries. From the application point of view, the main efforts are still devoted to investigate the superconducting properties, to distinguish intrinsic from extrinsic behaviours and to compare the different families in order to identify which one is the fittest for the quest for better and more practical superconductors. The 1111 family shows the highest $T_c$, huge but also the most anisotropic upper critical field and in-field, fan-shaped resistive transitions reminiscent of those of cuprates, while the 122 family is much less anisotropic with sharper resistive transitions as in low temperature superconductors, but with about half the $T_c$ of the 1111 compounds. An overview of the main superconducting properties relevant to applications will be presented. Upper critical field, electronic anisotropy parameter, intragranular and intergranular critical current density will be discussed and compared, where possible, across the Fe-based superconductor families.

## 1. Introduction

In 2008 the Hosono group in the Tokyo Institute of Technology discovered superconductivity at 26 K in the oxypnictide LaFeAs(O,F).[1] After only one month the critical temperature, $T_c$, doubled thanks to substitutions of the La by different rare earth (RE) elements (Sm, Ce, Nd, Pr and Gd) yielding an increase up to 55 K with Sm[2]. The parent compounds exhibit antiferromagnetic ordering of the iron moments which is suppressed by doping in favour of superconductivity. The early awareness that magnetic order, even if in competition with superconductivity, is a key factor for determining superconductivity, drove the discovery within a short period of new iron-based superconductor families with different crystal structures such as $(Ba,K)Fe_2As_2$,[3] LiFeAs,[4] and FeSe.[5] A large number of different compounds have now shown that superconductivity can be induced by carrier doping, both in the Fe-As layer and in the spacing layer, and by external as well as by internal pressure. For simplicity in the following we will refer to the different families as: 1111 (REFeAs(O,F)), 122 ((Ba,K)Fe$_2$As$_2$), 11(Fe(Se,Te)), 111(LiFeAs).

These four families share several characteristics with the cuprate superconductors, such as layered structure, the presence of competing orders, low carrier density, small coherence length, and unconventional pairing, all of which potentially hinder practical applications, especially due to their influence in inciting large thermal fluctuations and depressed grain boundary superconductivity. On the more positive side however, the Fe-based superconductors have metallic parent compounds, their anisotropy is generally smaller and does not strongly depend on the level of doping, and their generally supposed order parameter symmetry is s-wave, which is in principle not so detrimental to current transport across grain boundaries.

As in the early times of the cuprate superconductors, the main efforts are still devoted to distinguish intrinsic from extrinsic behaviour. The absence of significant transport currents in polycrystalline samples[6,7,8] has raised the question whether the low connectivity is an extrinsic effect due to low density, spurious phases, cracks, or an intrinsic depression of the superconducting order parameter similar to that observed in cuprates for more than very small angle grain boundary misorientations[9,10].

The availability of different pnictide families allows us to compare them and so to identify trends that might provide a clue for understanding the nature of superconductivity in these compounds, as well perhaps allowing us to focus on those matching the quest for better and more practical superconductors. The 1111 family, indeed, shows larger $T_c$, huge but also anisotropic upper critical field and in-field, fan-shaped resistive transition reminiscent of those of cuprates,[11,12] while the 122 family is less anisotropic and exhibits narrow resistive transitions like those in low temperature superconductors.[13,14]

In the following an overview of the principal superconducting properties relevant to applications is presented. In the first section, the upper critical field, $H_{c2}$, the electronic anisotropy, the coherence lengths, the paramagnetic limit and the effect of thermal fluctuations are discussed and compared across the Fe-based superconductor families. In the second section, the critical current densities of single crystals, polycrystals and bicrystals are reviewed.

**2. Upper Critical Fields**

The huge upper critical field values of Fe-based superconductors require investigation in high magnetic field laboratories. Already the first magnetoresistance measurements of polycrystalline La-1111 up to 45 T,[15] indicated a $\mu_0 H_{c2}$ value larger than 60 T which corresponds to a small coherence length of the order of few nm. Moreover, $H_{c2}(T)$ was anomalous and exceeded the Werthamer-Helfand-Hohenberg (WHH) formula [16], similar to that observed in dirty MgB$_2$ [17,18], suggesting that superconductivity in oxypnictides results from at least two bands. By replacing La with smaller rare earths like Nd and Sm, $T_c$ and $H_{c2}(0)$ both increase.[11] Going from the lower (La-1111) to the higher $T_c$ compounds (Nd-1111, Sm-1111), the in-field superconducting transitions become broader, approaching the broad magnetoresistive transitions of the cuprates for the highest $T_c$ compounds. The $H_{c2}$ slope at $T_c$ increases with increasing $T_c$, reaching a slope of 9.3 T/K in Sm-1111; even using WHH extrapolations which clearly underestimate many measurements, such $dH_{c2}/dT$ values yield $\mu_0 H_{c2}(0) \approx 0.693 T_c \mu_0 |dH_{c2}/dT|_{Tc} \approx 400 T$, much larger than the paramagnetic limit.

The availability of single crystals, first of the 1111 compounds, allows the evaluation of $H_{c2}$ parallel, $H_{c2}^{//ab}$, and perpendicular, $H_{c2}^{\perp ab}$, to the $ab$-plane.[12] The temperature dependence is very different in the two directions, strongly departing from the WHH behaviour[16] mainly in the direction parallel to $c$. The anisotropy evaluated as $\gamma = \gamma_H = H_{c2}^{//ab}/H_{c2}^{\perp ab}$, is also strongly temperature dependent, reminiscent of the two-gap behaviour seen in MgB$_2$.[17,18] However, a different situation is observed in the 122 family. (Ba,K)Fe$_2$As$_2$ single crystals exhibit nearly isotropic $\mu_0 H_{c2}$ with values of the order of 60 T at zero temperature, and anisotropy going from 2, close to $T_c$, down 1 at 5 K.[13] Similar results were reported in Ba(Fe,Co)$_2$As$_2$.[14]

These aspects have been investigated in high magnetic field on three single crystals belonging to the different families of the Fe-based superconductors. The main properties of these samples (NdFeAsO$_{0.7}$F$_{0.3}$ (Nd-1111 in the following), Ba(Fe$_{0.9}$Co$_{0.1}$)$_2$As$_2$ (Ba-122) and FeSe$_{0.5}$Te$_{0.5}$ (Fe-11)[19],[20],[21] with critical temperatures of 47.4, 22.0 and 14.5 K respectively defined at 50% of the normal state resistivity) are summarized in Table I.

|  | Nd-1111 | Ba-122 | Fe-11 |
|---|---|---|---|
| $T_c(50\%R_n)$ [K] | 47.4 | 22.0 | 14.5 |
| $\mu_0 dH_{c2}^{\perp ab}/dT$ [T/K] | 2.1 | 2.5 | 14 |
| $\mu_0 dH_{c2}^{//ab}/dT$ [T/K] | 10.1 | 4.9 | 26 |
| $\gamma_H$ | 5 | 1.9-1.5 | 1.9-1.1 |
| $\xi_{ab}$ [nm] | 1.8 | 2.4 | 1.2 |
| $\xi_c$ [nm] | 0.45 | 1.2 | 0.35 |
| Ginzburg number $G_i$ | $8\times10^{-3}$ | $1.7\times10^{-4}$ | $1.3\times10^{-3}$ |

Table 1: Significant superconducting state properties of pnictide single crystals.

Magneto-transport measurements were performed in a 16 T Quantum Design PPMS and in high magnetic field in the 35 T resistive and 45 T hybrid magnets at the National High Magnetic Field Laboratory (NHMFL). Figure 1 shows the temperature dependence of magnetoresistance of the three single crystal samples of Nd-1111, Ba-122 and Fe-11 in magnetic field applied parallel to *c*-axis. For Nd-1111, the transitions broaden with increasing the magnetic field, while for Ba-122 the breadth appears independent of field as in low temperature superconductors like Nb$_3$Sn. [22] For Fe-11 the situation is intermediate, even though actually this has a lower $T_c$, 14.5 K, as compared to 22 K for the 122 and 47 K for the 1111 single crystals.

Figure 2 shows $H_{c2}(T)$ in parallel and perpendicular field configurations determined with the 90% criterion. The three materials differ not only in $T_c$ and absolute values of $H_{c2}$ but importantly too in their temperature dependence of $H_{c2}$. Nd-1111 has a linear behavior in both directions whereas Ba-122 and Fe-11 show an almost linear behavior in the perpendicular direction but exhibit a downward curvature in the parallel direction. The slope of $H_{c2}$ close to $T_c$ significantly varies in the different families. $\mu_0 dH_{c2}^{\perp ab}/dT$ near $T_c$ varies from 2 T/K in Nd-1111 to almost 14 T/K in Fe-11 and $\mu_0 dH_{c2}^{//ab}/dT$ from 5 T/K in Ba-122 to the very high value of 25 T/K in Fe-11. The $H_{c2}$ anisotropy $\gamma_H$ is particularly affected by the different temperature dependences in the two directions. While the anisotropy is almost constant and equal to 5 in the Nd-1111, in the other two compounds it decreases with decreasing temperature. In Fe-11, for instance, the anisotropy close to $T_c$ is about 2 but, due to the downward curvature of the parallel direction, $\gamma_H$ approaches 1 at the lowest measured temperature.

*Paramagnetic limit*

The description of such upper critical field behavior is beyond the single-band, weak-coupling WHH model, where $H_{c2}$ is limited by orbital pair breaking, $\gamma_H$ is temperature independent and $dH_{c2}/dT$ is proportional to $T_c(1+\lambda)/v_F l$, where $\lambda$ is the electron-boson coupling constant, $v_F$ is the Fermi velocity and *l* is the electron mean free path. On the other hand the paramagnetic limit, where the superconductivity suppression is due to the alignment of the spins, is another pair-breaking mechanism to take into account.[23] In fact the upper critical field of those materials strongly exceeds the BCS paramagnetic limit, $H_p^{BCS}$ which is given by $\mu_0 H_p^{BCS}[T] = 1.84 T_c[K]$ [24] as emphasized in Figure 3 where the $\mu_0 H_{c2}/T_c$ as a function of $T/T_c$ is reported. Those factors suggest a more complex scenario. First of all to explain such a high $H_{c2}$, strong coupling has to be considered. The single-band Eliashberg theory allows the paramagnetic limit to be enhanced up to $\mu_0 H_P = 1.84 T_c (1+\lambda)$, strongly exceeding the BCS limit.[25] The effect of the paramagnetic limit may explain the different behavior of the samples considered here. In fact Nd-1111, whose $H_{c2}(T)$ in the experimentally accessible field range is well below the paramagnetic limit, shows a linear trend; as reported in Figure 3, while Ba-122 and Fe-11 reach higher values of $\mu_0 H_{c2}/T_c$ which exceed the BCS paramagnetic limit. In the case of Ba-122, paramagnetic suppression is mainly evident in the parallel direction where downward curvature is observed, while for Fe-11 both $H_{c2}$ orientations show downward curvature, suggesting that both are affected by Pauli pair breaking. Because paramagnetic limitation is isotropic, a stronger effect is expected in the parallel direction of higher $H_{c2}$, which should induce an anisotropy which reduces with decreasing temperature, as observed in Ba-122 and Fe-11. However a temperature dependent anisotropy may be explained also by multiband effects, as suggested in ref. 26 for $Sr(CoFe)_2As_2$ epitaxial film. A combination of the two mechanisms cannot be excluded.

*Fluctuation effects*

From the $H_{c2}$ slope close $T_c$ we may evaluate the in-plane, $\xi_a$, and out of plane, $\xi_c$, coherence lengths from the Ginzburg-Landau expressions $\xi_a = [\phi_0 / 2\pi\mu_0 (dH_{c2}^{//c}/T_c)T_c]^{1/2}$ and $\xi_c = \xi_a / \gamma_H$, as reported in Table 1. Interestingly, Fe-11 with the smallest $T_c$, presents the smallest coherence length values. More generally, the values are small for all samples and $\xi_c$ is comparable to the distance between the superconducting Fe-layers, as for the $CuO_2$ layers in the cuprate superconductors. It was suggested that thermal fluctuations may cause the broadening of the in-field resistive transition observed in the 1111 family[11] and a two dimensional fluctuation regime was indeed observed in magneto-conductance measurements of Sm-1111 compounds.[27] To understand whether fluctuation effects play a role also in the 122 and 11 families we evaluate the Ginzburg number $G_i$, which quantifies the temperature region $Gi \cdot T_c$, where fluctuation are significant. It is expressed by [11] $G_i = (\pi\lambda_0^2 k_B T_c \mu_0 / 2\xi_c \Phi_0^2)^2$ where $\lambda_0$ is the London penetration depth, $k_B$ is the Boltzmann constant and $\Phi_0$ is the flux quantum. Ginzburg numbers, evaluated for the three compounds assuming for simplicity $\lambda_0$=200 nm for all the compounds [14,28,29] are reported in table 1. The $G_i$ value obtained for Nd-1111 ($8 \times 10^{-3}$) is the largest and is comparable with the value obtained for YBCO ($10^{-2}$). The smallest number is obtained for Ba-122 ($1.7 \times 10^{-4}$), consistent with the narrow transitions in Fig. 1 and the low temperature superconductor-like behaviour emphasized in ref. 14. The number we obtain for Fe-11 ($1.3 \times 10^{-3}$), even if one order of magnitude lower than that obtained for high $T_c$ superconductors, is four orders of magnitude larger than the value estimated for a low $T_c$ superconductor with the same $T_c$ such as $V_3Si$. This makes the 11 family unique in being a low $T_c$ superconductor with an extremely short coherence length.

In order to detail the effect of fluctuations, resistivity measurements have been performed on an epitaxial film. The film, grown by pulsed laser deposition by a target of nominal composition $FeSe_{0.5}Te_{0.5}$ has $T_c$=18 K, larger than that of the target due to the strain developed during the growth[30]. The fluctuation conductivity $\Delta\sigma$ is evaluated as $\Delta\sigma = (\rho_n - \rho)/\rho\rho_n$ where $\rho_n$ is the normal state resistivity. In the inset of figure 4, $\rho$ and $\rho_n$ as linearly extrapolated in the range above $2T_c \approx 40K$ are shown. In the main panel of the same figure, the fluctuation conductivity is plotted versus $\varepsilon = \ln(T/T_c)$. We identify the so called Gaussian regime in the range $0.01 < \varepsilon < 0.1$, that is for temperatures from 0.4K to 2K above $T_c$, between the critical regime very close to $T_c$ and the high temperature regime of vanishing fluctuation conductivity. In this Gaussian regime, the behaviour of $\Delta\sigma$ is well described by the 3D law $\Delta\sigma = e^2/32\hbar\xi_c\sqrt{\varepsilon}$ where $\hbar = h/2\pi$ with h = Planck's constant (continuous line in figure 4). This 3D conclusion is consistent with the fact that $2\xi_c$ is of the order of the interplanar distance s=6.05Å. Indeed, the value of $\xi_c$ obtained by fitting fluctuation conductivity data is of the order of 1 nm, in agreement within a factor 2 with the value of 0.6nm extracted from the critical field data of this same film.

## 3. Critical current behaviour

Early studies of the critical current density ($J_c$) of 1111 polycrystalline samples emphasized the strong granularity of these compounds, which restricted global $J_c$ values to very low values.[6,7,8,31] A first optimistic claim came from Yamamoto et al.,[8] who found evidence for two distinct scales of current flow in polycrystalline Sm and Nd iron oxypnictides using magneto-optical imaging (MO) and study of the field dependence of the remanent magnetization. Such granular behaviour has so far limited the properties of pnictide wires,[32] even if wetting grain boundary phases and other extrinsic material inhomogeneities are one of the clear causes of this granularity. Even with substantial blocking by such grain boundary phases, the intergranular current densities appear to be more than one order of magnitude larger at 4 K than for early results on randomly oriented polycrystalline cuprates[33]. But certainly these early results make it clear that pnictides have different properties compared to randomly oriented $MgB_2$ polycrystals,[34] where grain boundaries can also partially obstruct without evidence for intrinsic obstruction of current flow as in the cuprates or as now appears to be the case in the pnictides.[35]

Before discussing the GB properties in the next section, we focus on bulk $J_c$ properties mainly obtained from single crystals and discuss the operating flux pinning mechanisms and the anisotropy of $J_c$.

*$J_c$ in single crystals*

As is usually the case, $J_c$ for single crystals must be evaluated by magnetization measurement and use of the Bean model, a procedure almost always possible for field $H$ parallel to the c-axis but much less frequently possible for $H$ parallel to ab-plane, where problems of the small size of crystals, significant anisotropy and difficulty of aligning crystals accurately with the field axis make extraction of $J_c$ from the measured magnetic moment uncertain. For the 1111 class, Zhigadlo et al. reported a high in-plane $J_c$ of $\sim 2\times 10^6$ A/cm$^2$ at 5 K on a $SmFeAsO_{1-x}F_x$ crystal. $J_c$ is almost field-independent up to 7 T at 5 K.[36] Many single crystal results were reported in the 122 system, since larger crystals can be easily grown. Yang et al. reported significant fishtail peak effects and large current carrying capability up to $5\times 10^6$ A/cm$^2$ at 4.2 K in a K-doped $Ba_{0.6}K_{0.4}Fe_2As_2$ single crystal.[37] Yamamoto et al. deduced $J_c \sim 4\times 10^5$ A/cm$^2$ at 4.2 K and also reported the fishtail peaks in their Co-doped $Ba(Fe_{0.9}Co_{0.1})_2As_2$ single crystal.[14] Prozorov et al. showed $J_c$ of $2.6\times 10^5$ A/cm$^2$ at 5 K for $Ba(Fe_{0.93}Co_{0.07})_2As_2$ single crystals and also showed fishtail peaks as well as very large magnetic relaxation rate, which were analyzed using collective pinning and creep models.[38] As for the 11 system, Taen et al. reported that $J_c$ of tellurium doped $FeTe_{0.61}Se_{0.39}$ crystals with $T_c \sim 14$ K exceeded $1\times 10^5$ A/cm$^2$ at low temperatures.[39]

All these results show that Fe-based superconductors exhibit rather high $J_c$ values, independent of the field at low temperatures similar to the behavior observed in YBCO.[40] Such results are all consistent with the nm-scale coherence lengths in Table I, the exceptionally high $H_{c2}$ values and pinning associated with atomic-scale defects, resulting from chemical doping. The common fishtail observation may indicate the presence of nanoscale phase separation into regions of weaker superconductivity that are proximity-coupled to the higher-$T_c$ matrix, perhaps an intrinsic effect or one caused by an inhomogeneous distribution of the Co or K doping agent. Irradiation with Au-ions and neutrons has emphasized that pinning can be further increased by introducing defects without affecting $T_c$. Au-ions produce columnar defects that increase the critical current density, but less than one order of magnitude at low field.[41] Similar results were obtained with neutron irradiation which produces a more isotropic defect structure.[42] In this respect too the pnictides appear quite similar to the cuprates.

Because large single crystals of 122 can be grown, it is possible to study their anisotropy magnetically. In a recent study, single crystals $0.16 \times 0.93 \times 1.3$ mm$^3$ of Ba(Fe$_{0.9}$Co$_{0.1}$)$_2$As$_2$ with a sharp $T_c$ transition of 23 K were grown at the NHMFL using the FeAs flux method. Magnetic fields of up to 14 T were applied both parallel to the *c*-axis and *ab*-plane of the crystal in an Oxford vibrating sample magnetometer (VSM). Figures 5 (a) and (b) show magnetic hysteresis loops at 4.2-20 K in both orientations. All loops show negligible background ferromagnetic moment, indicative of little free Fe, often present in such crystals. Loops for both configurations show large hysteresis and a slight fishtail in *M*(*H*), consistent with strong pinning.

Extraction of the anisotropic $J_c$ depends on assumptions about the anisotropic Bean model and current scale that we assume to be the full sample size. For *H*//*c*, it is reasonable to assume that the currents flow in the *ab*-plane and the Lorentz force drives vortices perpendicular to the *ab*-planes. We call this current $J_{c,ab}$. If *H*//*a*(*b*) the currents flow along *b*(*a*) and *c*; since our crystal is a platelet whose size normal to the *ab*-planes is smaller than along the *ab*-planes, the main hysteretic moment comes from currents flowing in the *b*(*a*) direction and the Lorentz force driving vortices along the *c*-axis. We call this current $J_{c,c}$.

The field dependence of $J_{c,ab}$ was calculated on the basis of the Bean model and is shown in Figure 6. The value of $J_{c,ab}$ is $5.3 \times 10^5$ A/cm$^2$ at 4.2 K, which is similar to that reported previously.[14,38,41] The field dependence of $J_c$ is rather mild, especially at low temperatures, consistent with the high $H_{c2}$. In order to obtain the anisotropy of $J_c$, we deduced $J_c$ along *c*-axis ($J_{c,c}$) from hysteresis loop in field parallel to *ab*-plane using the extended Bean model. It was assumed that $J_{c,ab}$ under self-field does not change much regardless of *H*//*ab* and *H*//*c*. The estimated $J_{c,c}$ at 4.2 K under self-field is $\sim 1.3 \times 10^5$ A/cm$^2$. The temperature dependence of $J_{c,ab}$ and $J_{c,c}$ are fitted well with an expression $J_c = J_c(0) \times (1-T/T_c)^n$ with $J_{cab}(0) = 7.5 \times 10^5$ A/cm$^2$, $n = 1.75$ for $J_{c,ab}$ and $J_{c,c}(0) = 2.0 \times 10^5$ A/cm$^2$, $n = 2$ and shown in Fig. 7. In the inset of figure 7 the temperature-dependent anisotropy of $J_c$ is plotted. The anisotropy $\gamma_J = J_{c,ab}/J_{c,c}$ is $\sim 6$ near $T_c$, decreases with decreasing $T$ and reaches $\sim 4$ at low temperatures. The obtained value of $\gamma_J$ is consistent with the value $\gamma_J \sim 2-3$ reported by Tanatar *et al.*[43]

Analysis of the pinning force curve gives us insight into the underlying vortex pinning mechanisms. It is well known that the pinning force $F_p = J_c \times \mu_0 H$ of conventional metallic superconductors scales as $F_p \sim H_{c2}^m h^p (1-h)^q$, where $h = H/H_{irr}$ is the ratio between $H$ and the irreversibility field $H_{irr}$. Here we show that the normalized pinning force $f_p = F_P/F_P^{max}$ as a function of reduced field $h$ obtained from hysteresis loops in field applied parallel to *c*-axis (Figure 8 upper panel) and *ab*-plane (Figure 8 lower panel); $H_{irr}$ has been estimated from a Kramer plot.[44] $H_{irr}$ values evaluated in the two directions differ by a factor 2, consistent with the $H_{c2}$ anisotropy. As discussed above, $F_p(H//c)$ and $F_p(H//ab)$ are considered to be mainly determined by critical currents flowing in the *ab*-plane and vortex motion along planes and across planes, respectively. Pinning force curves for both parallel and perpendicular field configurations scale well independently of temperature (from 4.2 to 17.5 K). This suggests that a single dominant vortex pinning mechanism works at all temperatures. The pinning force curves collapse according to the law $f_p \propto h^p(1-h)^q$ with $p = 1.1$ and $q = 3$ for *H*//*c* and $p = 1.25$ and $q = 2.25$ for *H*//*ab*, respectively, as plotted in dashed lines in Figures 8. Maximum of $f_p(h)$ curves occur at $h \sim 0.25$ for *H*//*c*, $h \sim 0.35$ for *H*//*ab*. For a Ba$_{0.6}$K$_{0.4}$Fe$_2$As$_2$ single crystal the maximum is found at $h=0.33$.[37] The peak position may give clues to the pinning mechanism. As a first approximation we can assume that the position of the peak shifts to higher $h$ values with decreasing distance between pinning centers. Thus, once the $H_{c2}$ anisotropy is taken into account by scaling the data with $H_{irr}$, still a difference between $J_{c,ab}$ and $J_{c,c}$ survives. We can assume that for *H*//*c* localized defects pin vortices, while for *H*//*ab* the modulation of the order parameter along the *c*-axis could play a role. This is compatible with strong bulk pinning suggested from scanning tunneling spectroscopy.[45] Recently, in single crystals of Ba(Fe$_{1-x}$Co$_x$)$_2$As$_2$ a combination of polarized-light imaging and magnetic measurements have show that the pinning is significantly enhanced by orthorhombic magnetic/structural domains.[46] This mechanisms is supposed to be intrinsic to this phase and more significant in the slightly underdoped compositions.

*Global $J_c$ and Grain Boundary effects in polycrystalline materials*

Practical use of the pnictides in large scale applications would be greatly enhanced if polycrystalline forms were not intrinsically electromagnetically granular, as is the case for the cuprates. That the pnictides are granular has been raised by multiple studies of polycrystals in bulk forms,[6,7,8] in wire forms,[32] and also in thin film forms.[47,48,49,30] But whereas it was relatively easy to get single-phase polycrystalline forms of the cuprates, it appears to be much harder in the case of the pnictides. We here briefly review recent studies of current transport in polycrystalline Sm- and Nd-1111[8,50,51] that were made by high pressure synthesis at the Institute of Physics in Beijing (IOP-CAS). We have benchmarked these IOP-CAS samples of Ren *et al.* against carefully made Sm-1111 samples made at the NHMFL and in INFM-LAMIA in Genoa, with and without benefit of hot isostatic pressing. We find the intergranular connectivity of the IOP-CAS Sm-1111 sample to be the highest of all, even though there is clear evidence of significant wetting FeAs phase and unreacted RE$_2$O$_3$, impurity phases found to be common to all. Thus we believe that these results have substantial general validity. Moreover, we find that polycrystalline samples of Co-doped 122 have 50-100 μm diameter grains, rather than the 5-10 μm diameter grains in the 1111 polycrystals. Magneto-optical images show essentially complete decoupling across the grain boundaries, but also substantial FeAs phase that wets the grain boundaries, as shown in Figure 9. Only in the recent work of Lee *et al.*[35] does the intrinsic behavior of second-phase free grain boundaries appear. Unfortunately it appears that symmetric [001] tilt grain boundaries grown epitaxially on SrTiO$_3$ exhibit substantial depression of $J_c$ for misorientations of more than 3°.

The use of the low temperature laser scanning microscope (LTLSM) enables a direct spatial correlation between the position at which an electric field *E* occurs in the superconducting state and the microstructure with a precision of 1-2 μm. Figure 10 shows details of such correlations for two types of regions, type A and B that show dissipative super current flow only in self or very weak field up to ~0.1 T and region C where flow remains dissipative even in 5 T after regions A and B have switched off. The SEM images of Fig. 10 show significant microstructural differences between regions A and B, and C. Precipitates of unreacted Sm$_2$O$_3$ are the most benign current-blocking defects because, although insulating, they have a small surface to volume ratio, and mostly occur within Sm1111 grains. By contrast, the dark grey FeAs phase wets many grain boundaries, thus interrupting grain-to-grain supercurrent paths, which are further degraded by extensive cracking, sometimes at grain boundaries (the black-appearing lines) and sometimes within grains.

At switch-off spot A of Fig. 10 (a), a crack F on the upper side and the precipitate of Sm$_2$O$_3$ and a large FeAs phase G force current to cross grain boundary (H) containing a thin FeAs layer, producing the dissipation spot seen in the overlay image of Fig. 10 (d). At switch off point B of Fig. 10 (b) and (e), the current is channeled by cracks, Fe-As and Sm$_2$O$_3$ into a narrow passage crossing FeAs regions too. By contrast, as shown in Fig. 10 (c) and (f), spot C that remains dissipative even in 5 T field has its peak dissipation within a single grain at a constriction provided by two almost orthogonal sets of cracks which squeeze the current between the two diagonal cracks. The S-N-S (superconducting-normal-superconducting) nature of the connection across the metallic FeAs phase is strongly suggested by the very strong (10 fold) fall off of $J_c$ in even 0.1 T field. Detailed analysis by MO imaging and remanent field analysis of subdivided samples had earlier shown that the intergranular current was both much smaller (~4000 A/cm$^2$ at 4 K) and had an SNS-like temperature dependence, while the intra-granular current density significantly exceeded 10$^6$ A/cm$^2$

(see Figure 11). The reasonable conclusion to draw from these and many other studies of polycrystals is that granular behavior is quite evident but that one source of the granularity is uncontrolled second phase, particularly residual FeAs phase. Use of techniques such as remanent magnetization analysis, MO imaging and LTLSM imaging enable a quite detailed understanding of these effects.[8] The intragranular $J_c$ values are largely consistent with the results obtained from single crystal studies discussed above. All suggest that the pnictides are inherent nanomaterials because of their short coherence lengths and thus produce high densities of pinning defects. We may conclude from the rapid fall off of $J_c$ in Fig. 11(b) however that many of these are point defects that are easily thermally depinned at higher temperatures.

*$J_c$ in Thin Films*

Thin films have generally had significantly lower $J_c$ values than bulk single crystals and indeed have led to an independent conclusion that polycrystalline films exhibit electromagnetically granular behavior.

Thin films have not been easy to grow, especially of the 1111 compounds where doping is largely produced by F and by O and where both are volatile and effectively uncontrolled in the final films. Study of La1111 by the Dresden group has concluded that the largely polycrystalline forms produced by *ex situ* growth produces an electromagnetically granular film, even though single crystal LSAT substrates are used. In principle it should be possible to dope more easily in the Co-122 systems where the doping agent (Co) is not volatile and indeed this allows *in situ* growth and greater degrees of epitaxy [48,52,53]. However growth of the 122 structure on LSAT does not seem to produce genuine epitaxy and $J_c$ values are lower than those seen in bulk single crystals. Only in the films recently grown by the Eom group on $SrTiO_3$ or on LAO with STO intermediate layers is genuine epitaxy obtained[35, 54]. In this case $J_c$ values exceed $10^6$ A/cm$^2$ and films appear to be quite free of electromagnetic and crystalline granularity. Studies of the angular dependence of $J_c$ in these films also show significantly higher $J_c$ for $H$ parallel to $c$ than for $H$ parallel to $ab$, a result that contradicts the $H_{c2}$ anisotropy of the films and demonstrates that conventional strong vortex pinning effects are possible in the pnictides.[54]

Epitaxial growth on STO immediately suggests that the classic bicrystal experiment[55] is possible and indeed this has now been reported by Lee *et al*.[35] The key result is that $J_c$ is reduced by grain boundaries with 5-24° [001] tilt. Study of 3°, 6°, 9° and 24° bicrystals shows that there is a progressive reduction of $J_c$ with increasing misorientation that is reminiscent of, but not as strong as in cuprate, especially planar YBCO grain boundaries.

**4. Summary**

We have summarized recent studies of the pnictides from the viewpoint of potential applications. A key point is that they have properties intermediate between the LTS materials like Nb-Ti and Nb$_3$Sn and the cuprates like YBCO and Bi-2212 or Bi-2223. On their positive side is that they can have $T_c$ up to 55 K and $H_{c2}(0)$ well over 100 T.

After a comparison among the families 122 comes out the most suitable for application with rather high T$_c$, upper critical field, low anisotropy, reduced thermal fluctuations and intrinsic pinning mechanisms. In particular the Co-doped 122 compound with $T_c$ of ~22 K $H_{c2}(0)$ of >50 T, has almost twice that of Nb$_3$Sn (30 T) with a $T_c$ of 18 K. Although the Nb-base materials are isotropic, Co-122 is almost isotropic ($\gamma < 2$) too, making it potentially competitive as a low temperature superconductor. Even the highest $T_c$ pnictides, Sm- and Nd-1111 have anisotropies much smaller than typical cuprates ($\gamma \sim 30$). However a typical YBCO has $\gamma \sim 5$, similar to the 1111. A clear drawback to present applications of the pnictides is their extrinsic and perhaps intrinsic granularity that significantly restrict the critical current density of polycrystalline forms. However, since only 18 months have passed since the first reports of $T_c$ above 20 K in the pnictides, we should not expect that discoveries are yet over or that the final word on applications can yet be given.

**Acknowledgements**
A portion of this work was performed at the National High Magnetic Field Laboratory, which is supported by NSF Cooperative Agreement No. DMR-0654118, by the State of Florida, and by the US Dept. Of Energy. Explicit support for the pnictide work at the NHMFL comes from AFOSR under grant FA9550-06-1-0474. This work was also partially supported by the Italian Foreign Affairs Ministry (MAE) - General Direction for the Cultural Promotion. One of the authors (AY) is supported by a fellowship of the Japan Society for the Promotion of Science and (MP) by CNR under the project Short Term Mobility. The research at ORNL was sponsored by the Division of Materials Sciences and Engineering, Office of Basic Energy Sciences, U.S. Department of Energy.

**Figure captions**

Figure 1.
Magneto-transport measurements in high magnetic field applied perpendicularly to the samples for three different materials: $NdFeAsO_{0.7}F_{0.3}$ (Nd-1111), $Ba(Fe_{0.9}Co_{0.1})_2As_2$ (Ba-122) and $FeSe_{0.5}Te_{0.5}$ (Fe-11).

Figure 2.
Temperature dependences of $H_{c2}^{//ab}$ (filled symbols), and $H_{c2}^{\perp ab}$ (empty symbols) for the same samples of Figure 1. In the inset a magnification of the region close to $T_c$ for the Fe-11 sample.

Figure 3.
$\mu_0 H_{c2}/T_c$ as a function of $T/T_c$ Nd-1111, Ba-1222 and Fe-11 for H//ab (filled symbols) and H$\perp$ab (empty symbols). The broken line represent the BCS paramagnetic limit $\mu_0 H_P = 1.84 T_c (1+\lambda)$. In the inset a magnification of the region close to $T/T_c=1$

Figure 4.
Fluctuation conductivity $\Delta\sigma$ as a function of $\varepsilon=\ln(T/T_c)$ in a Log-Log scale: filled symbols represent the experimental data and continuous line represents the 3D Aslamazov-Larkin behaviour $\Delta\sigma = e^2/32\hbar\xi_c\sqrt{\varepsilon}$. Inset: low temperature resistivity data (open square symbols) and linear extrapolation of normal state resistivity data (continuous line).

Figure 5.
Magnetic hysteresis loops of the $Ba(Fe_{0.9}Co_{0.1})_2As_2$ crystal in field parallel to *c*-axis (a) and *ab*-plane (b) at 4.2, 7.5, 10, 12.5, 15, 17.5 and 20 K.,

Figure 6.
Critical current density for H//c ($J_{c,ab}$) as a function of magnetic field at 4.2÷20 K for the $Ba(Fe_{0.9}Co_{0.1})_2As_2$ crystal.

Figure 7.
Temperature dependence of critical current density along *ab*-plane ($J_{c,ab}$) and *c*-axis ($J_{c,c}$) for the $Ba(Fe_{0.9}Co_{0.1})_2As_2$ crystal. The dashed lines show fitting of the experimental data with $J_c =$

$J_c(0) \times (1-T/T_c)^n$ with $J_{c,ab}(0) = 7.5 \times 10^5$ A/cm$^2$, $n = 1.75$ for $J_{c, ab}$ and $J_{c, c}(0) = 2.0 \times 10^5$ A/cm$^2$, $n = 2$. Inset: Temperature dependence of anisotropy of the critical current density $\gamma_J = J_{c,ab} / J_{c, c}$.

Figure 8.
Normalized flux pinning force $f_p$ as a function of reduced field $h = H/H_{irr}$ of the Ba(Fe$_{0.9}$Co$_{0.1}$)$_2$As$_2$ crystal for field applied parallel to *c*-axis (a) and *ab*-plane (b).

Figure 9.
Optical image (left) of a polycrystal Ba(Fe$_{0.9}$Co$_{0.1}$)$_2$As$_2$ bulk sample and the corresponding magneto-optical image (right) taken after zero-field cooled and applied field of 100 mT at 6.4 K.

Figure 10.
(a)(b)(c) High magnification SEM images of a Sm1111 polycrystal that was polished down to 20 µm thickness and then examined in a low temperature laser scanning microscope in the superconducting state[51]. About 20 regions showing supercurrent dissipations were seen of which about ¾ switched off when more than ~0.1 T was applied. This image set shows microstructural details of typical switch-off spots A, B, and dissipation spot C that was still passing supercurrent at 5 T. There are second phases of Sm$_2$O$_3$ and FeAs with white and dark gray contrast in addition to platey Sm1111 grains. Cracks with dark line contrasts are also seen. (d)(e)(f) Dissipation spots at self field superimposed on the SEM images of (a)(b)(c), respectively. Deeper red colour represents the areas with stronger dissipation where higher supercurrent density was focused.

Figure 11.
(a): Temperature dependence of global critical current density $J_c^{global}(T)$ for the polycrystalline SmFeAsO$_{0.85}$ and NdFeAsO$_{0.94}$F$_{0.06}$ bulk samples obtained from the remanent magnetization analysis (filled), magneto optical $B(x)$ flux profile analysis[8]. (b) Temperature dependence of critical current density of locally circulating current $J_c^{local}(T)$ for the polycrystalline SmFeAsO$_{0.85}$ and NdFeAsO$_{0.94}$F$_{0.06}$ bulk samples obtained from remanent magnetization analysis. Inset shows log-scale plots for the SmFeAsO$_{0.85}$ experimental data with an exponential and linear fitting.

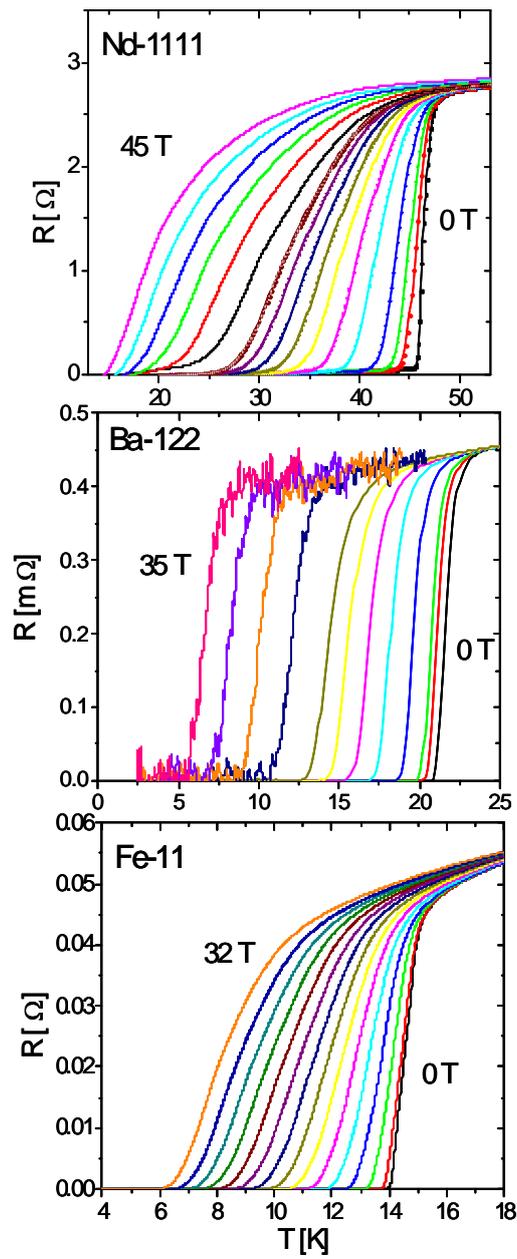

Figure 1.
Magneto-transport measurements in high magnetic field applied perpendicularly to the samples for three different materials: NdFeAsO$_{0.7}$F$_{0.3}$ (Nd-1111), Ba(Fe$_{0.9}$Co$_{0.1}$)$_2$As$_2$ (Ba-122) and FeSe$_{0.5}$Te$_{0.5}$ (Fe-11).

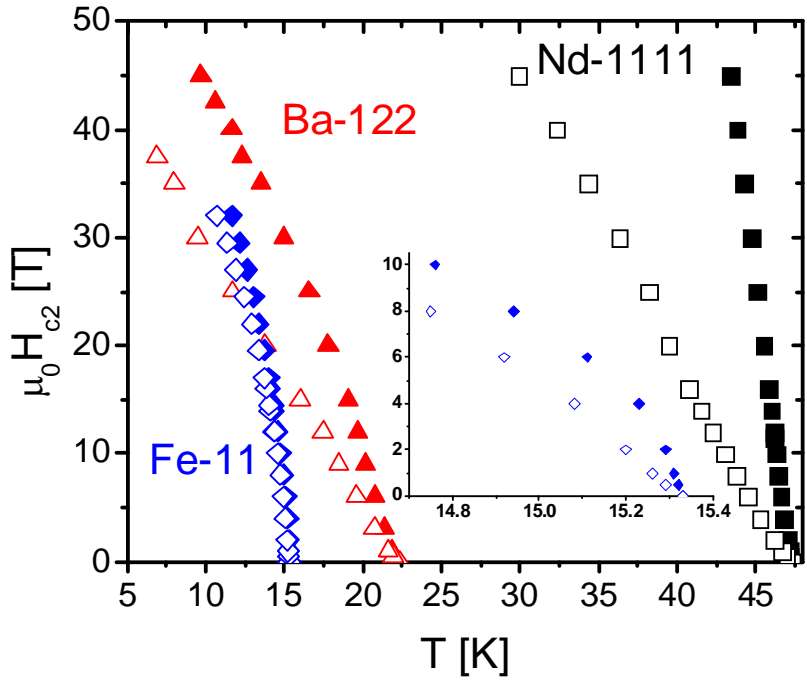

Figure 2.
Temperature dependences of $H_{c2}^{//ab}$ (filled symbols), and $H_{c2}^{\perp ab}$ (empty symbols) for the same samples of Figure 1. In the inset a magnification of the region close to $T_c$ for the Fe-11 sample.

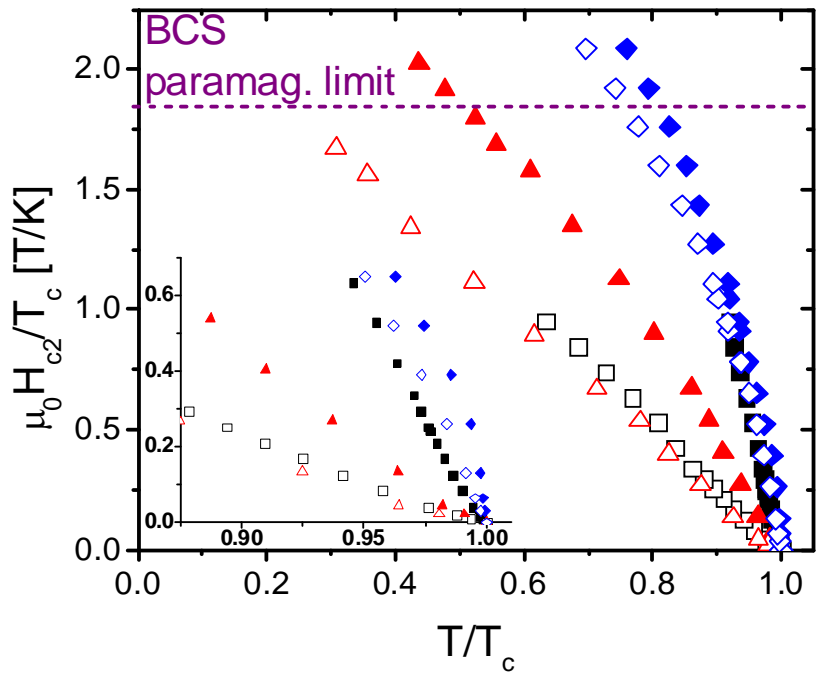

Figure 3.
$\mu_0 H_{c2}/T_c$ as a function of $T/T_c$ Nd-1111, Ba-1222 and Fe-11 for H//ab (filled symbols) and H⊥ab (empty symbols). The broken line represent the BCS paramagnetic limit $\mu_0 H_P = 1.84 T_c(1+\lambda)$. In the inset a magnification of the region close to $T/T_c=1$

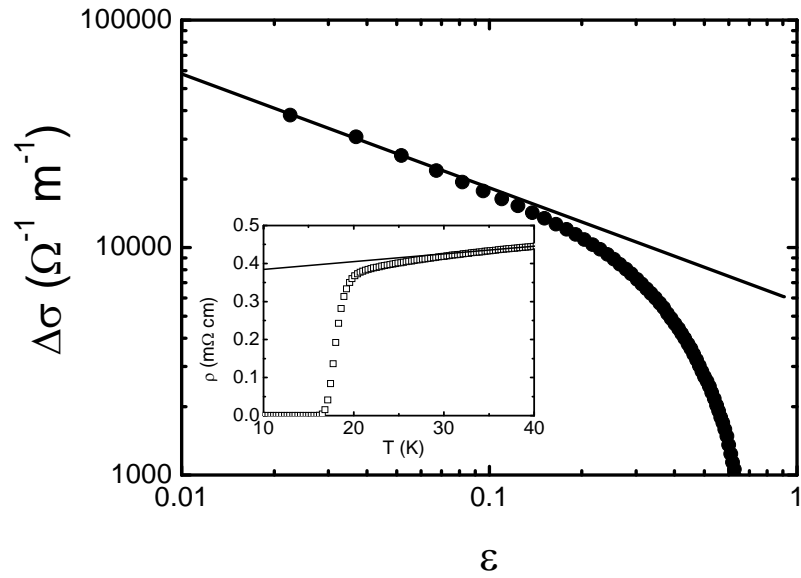

Figure 4.
Fluctuation conductivity $\Delta\sigma$ as a function of $\varepsilon=\ln(T/T_c)$ in a Log-Log scale: filled symbols represent the experimental data and continuous line represents the 3D Aslamazov-Larkin behaviour $\Delta\sigma = e^2/32\hbar\xi_c\sqrt{\varepsilon}$. Inset: low temperature resistivity data (open square symbols) and linear extrapolation of normal state resistivity data (continuous line).

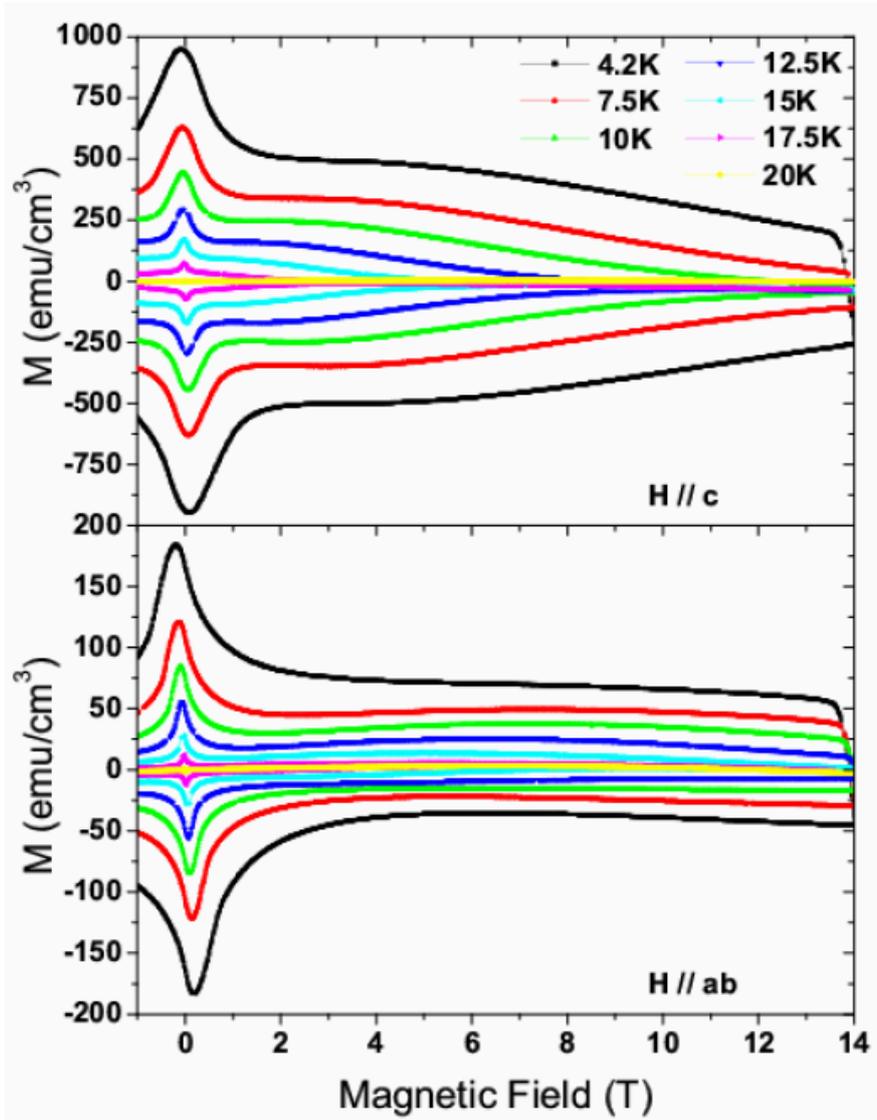

Figure 5.
Magnetic hysteresis loops of the Ba(Fe$_{0.9}$Co$_{0.1}$)$_2$As$_2$ crystal in field parallel to *c*-axis (a) and *ab*-plane (b) at 4.2, 7.5, 10, 12.5, 15, 17.5 and 20 K.

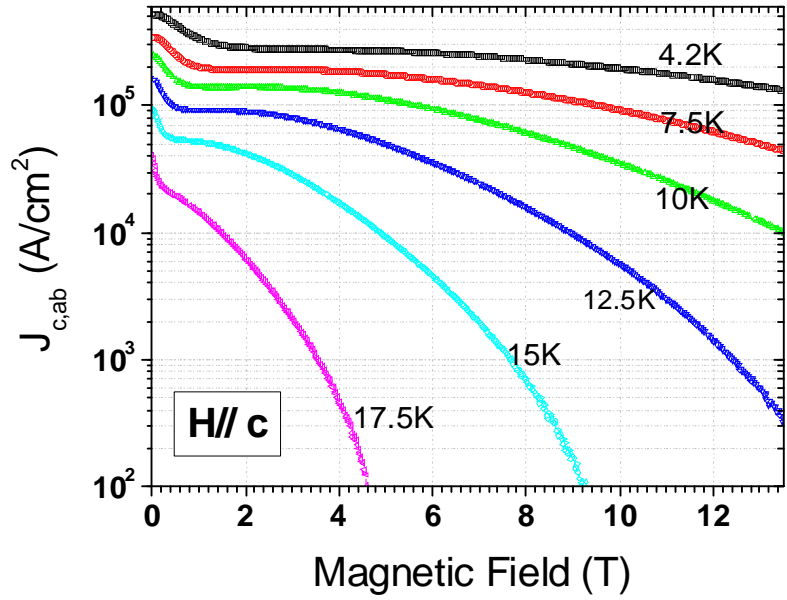

Fig. 6.
Critical current density for $H//c$ ($J_{c,ab}$) as a function of magnetic field at 4.2÷20 K for the Ba(Fe$_{0.9}$Co$_{0.1}$)$_2$As$_2$ crystal.

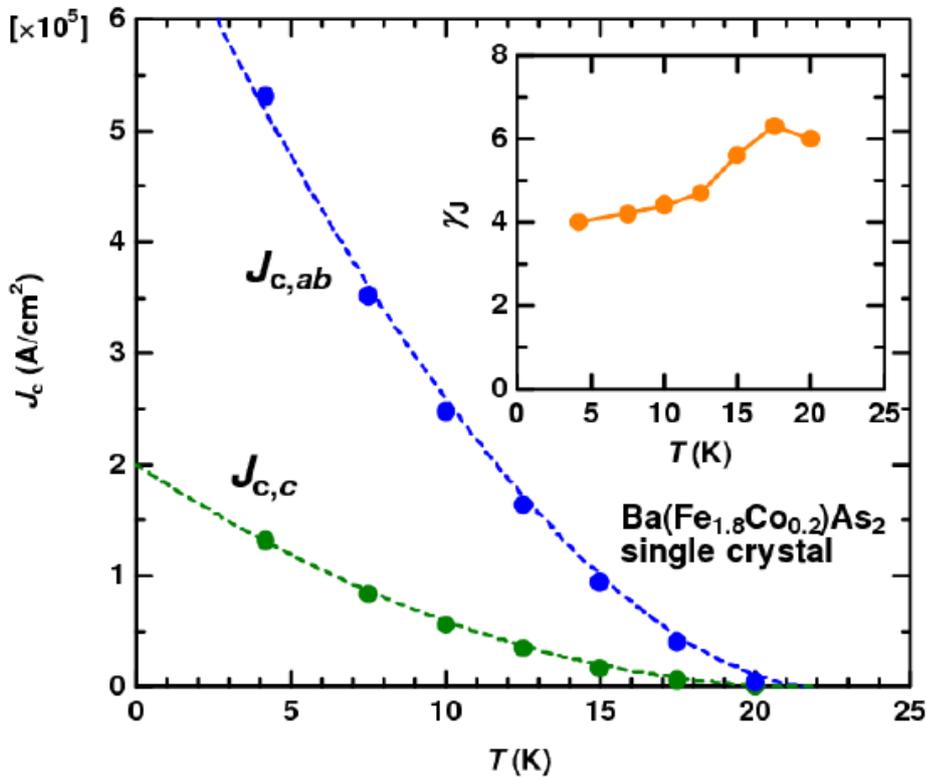

Fig. 7
Temperature dependence of critical current density along *ab*-plane ($J_{c,ab}$) and *c*-axis ($J_{c,c}$) for the Ba(Fe$_{0.9}$Co$_{0.1}$)$_2$As$_2$ crystal. The dashed lines show fitting of the experimental data with $J_c = J_c(0) \times (1-T/T_c)^n$ with $J_{c,ab}(0) = 7.5 \times 10^5$ A/cm$^2$, $n = 1.75$ for $J_{c,ab}$ and $J_{c,c}(0) = 2.0 \times 10^5$ A/cm$^2$, $n = 2$. Inset: Temperature dependence of anisotropy of the critical current density $\gamma_J = J_{c,ab} / J_{c,c}$.

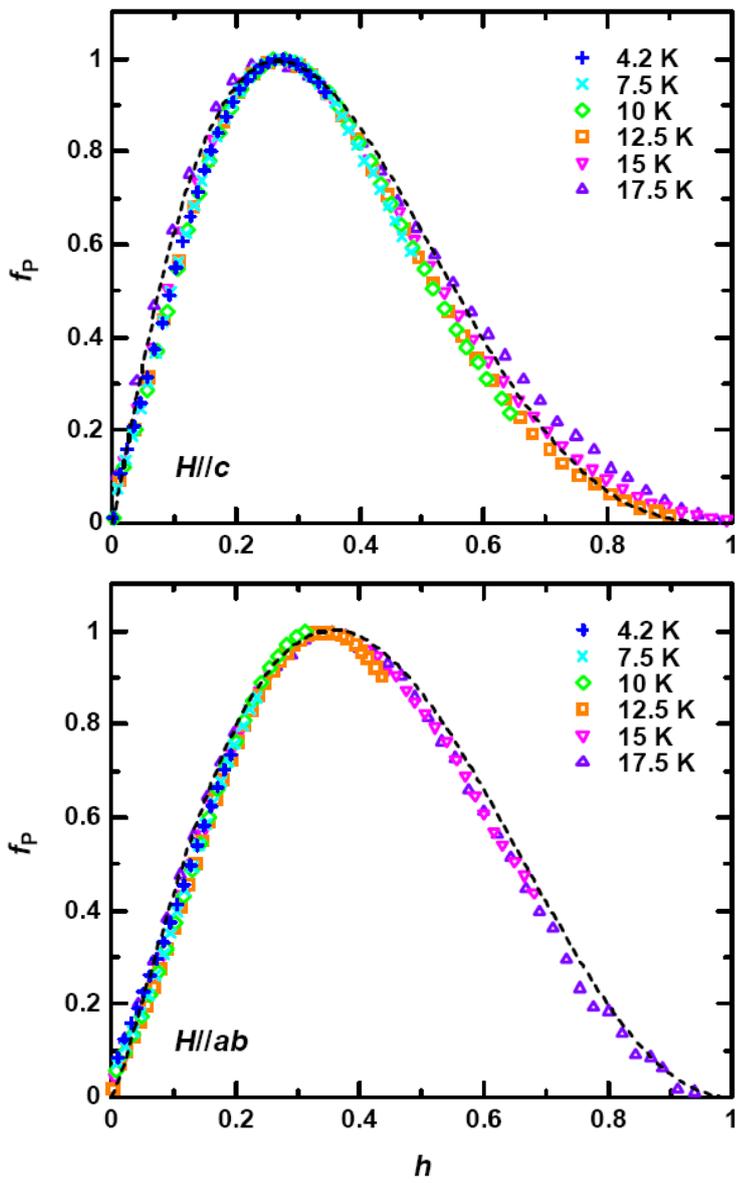

Fig. 8
Normalized flux pinning force $f_p$ as a function of reduced field $h = H/H_{irr}$ of the Ba(Fe$_{0.9}$Co$_{0.1}$)$_2$As$_2$ crystal for field applied parallel to c-axis (a) and ab-plane (b).

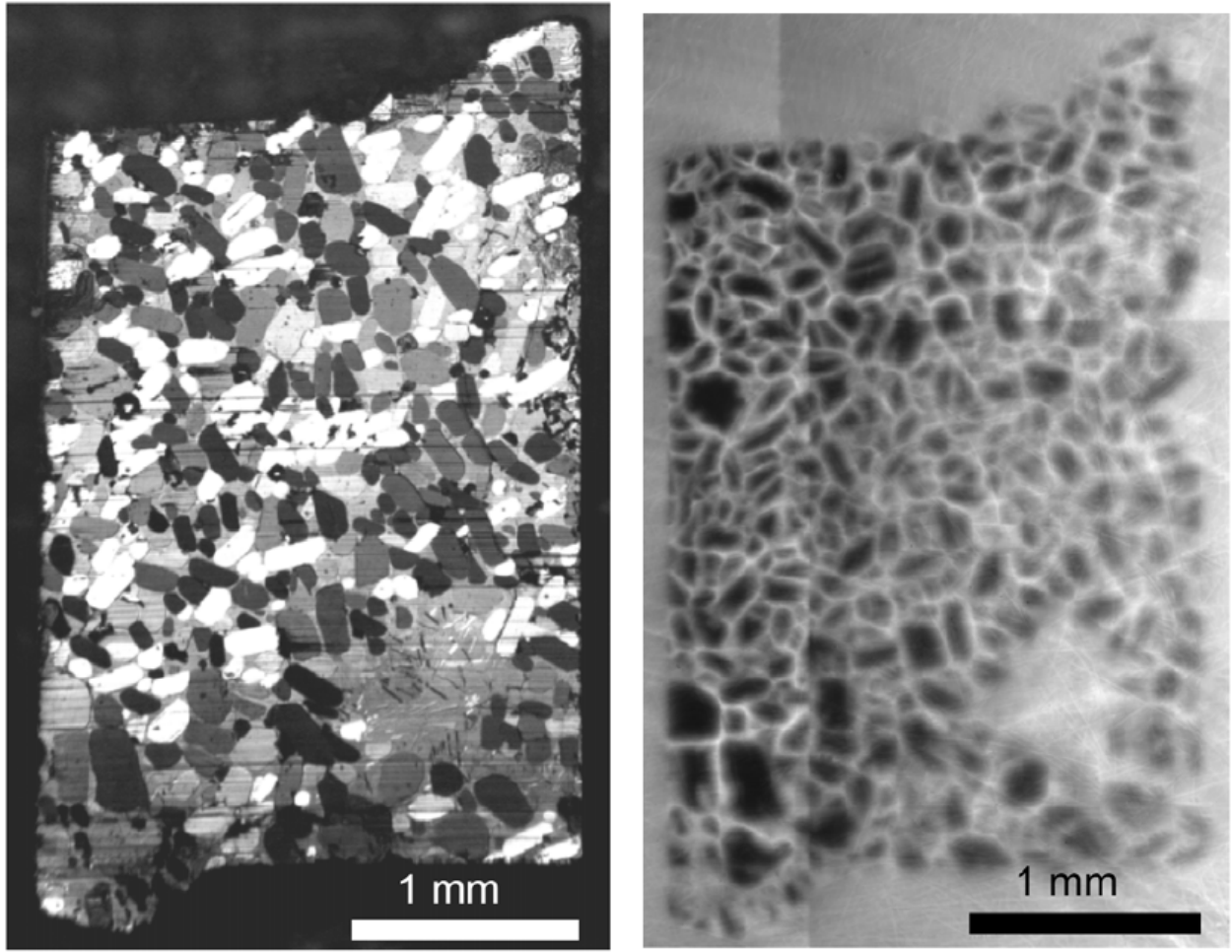

Fig. 9 Optical image (left) of a polycrystal Ba(Fe$_{0.9}$Co$_{0.1}$)$_2$As$_2$ bulk sample and the corresponding magneto-optical image (right) taken after zero-field cooled and applied field of 100 mT at 6.4 K.

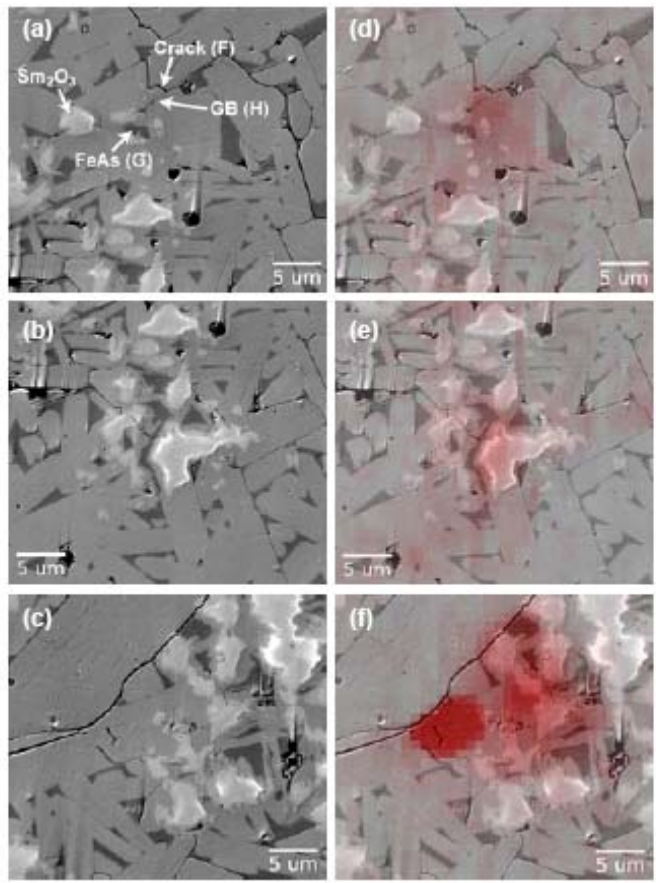

Fig. 10 (a)(b)(c) High magnification SEM images of a Sm1111 polycrystal that was polished down to 20 μm thickness and then examined in a low temperature laser scanning microscope in the superconducting state[51]. About 20 regions showing supercurrent dissipations were seen of which about ¾ switched off when more than ~0.1 T was applied. This image set shows microstructural details of typical switch-off spots A, B, and dissipation spot C that was still passing supercurrent at 5 T. There are second phases of $Sm_2O_3$ and FeAs with white and dark gray contrast in addition to platey Sm1111 grains. Cracks with dark line contrasts are also seen. (d)(e)(f) Dissipation spots at self field superimposed on the SEM images of (a)(b)(c), respectively. Deeper red colour represents the areas with stronger dissipation where higher supercurrent density was focused.

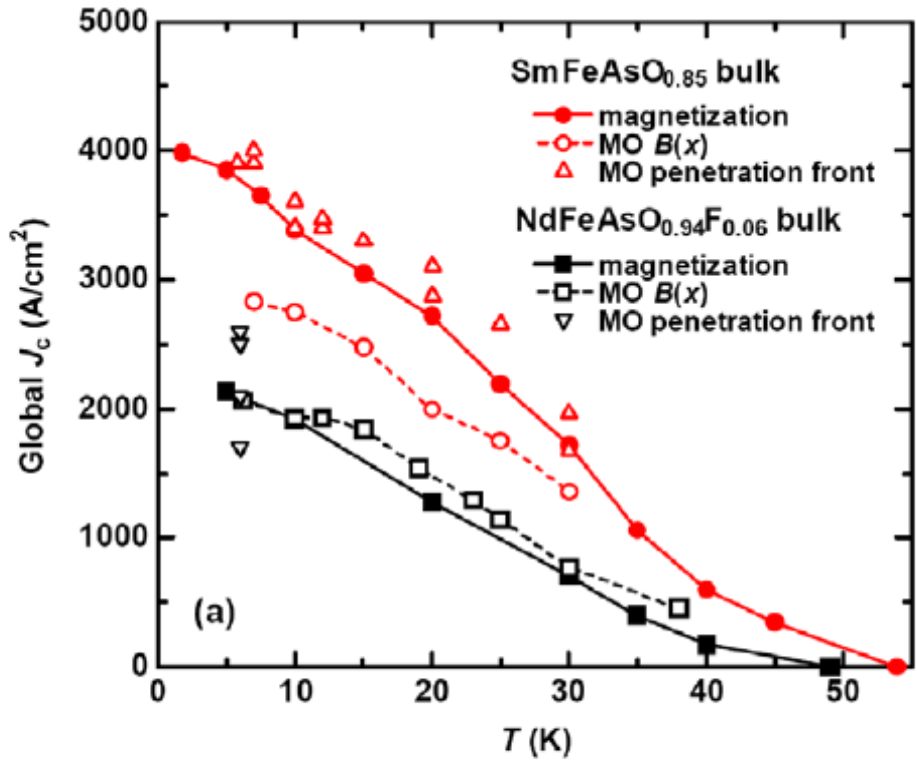

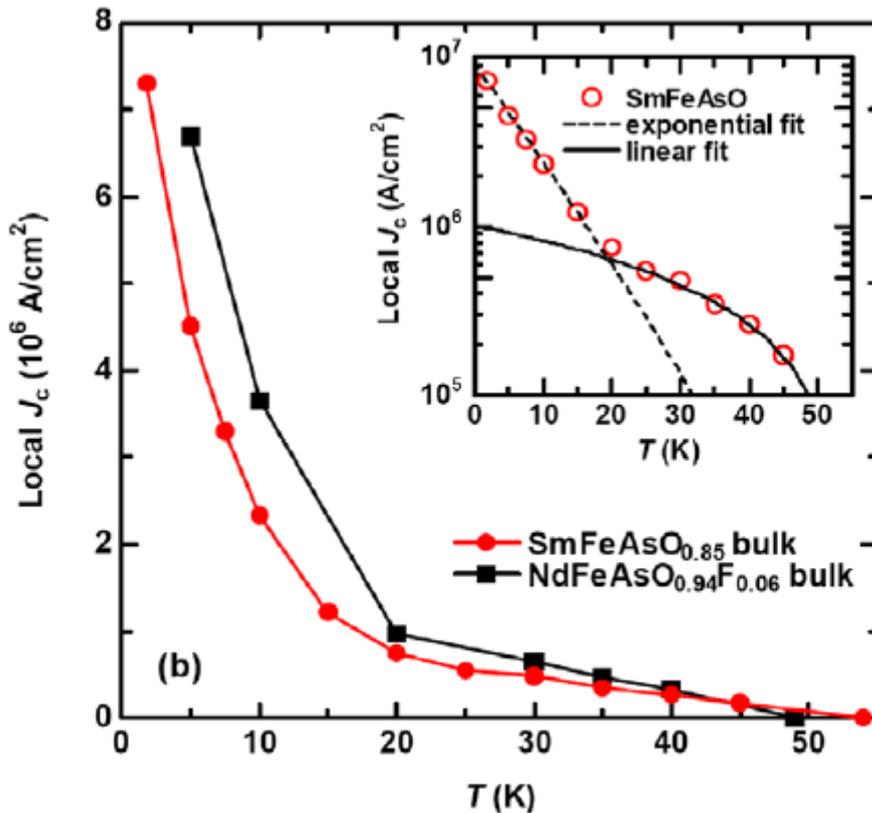

Fig. 11 (a): Temperature dependence of global critical current density $J_c^{global}(T)$ for the polycrystalline $SmFeAsO_{0.85}$ and $NdFeAsO_{0.94}F_{0.06}$ bulk samples obtained from the remanent magnetization analysis (filled), magneto optical $B(x)$ flux profile analysis[8]. (b) Temperature dependence of critical current density of locally circulating current $J_c^{local}(T)$ for the polycrystalline $SmFeAsO_{0.85}$ and $NdFeAsO_{0.94}F_{0.06}$ bulk samples obtained from remanent magnetization analysis. Inset shows log-scale plots for the $SmFeAsO_{0.85}$ experimental data with an exponential and linear fitting.